\documentclass[journal]{IEEEtran}

\ifCLASSINFOpdf
\else
\fi

\usepackage[dvips]{graphicx}
\usepackage{float}
\usepackage[font=footnotesize]{subfig}
\usepackage[cmex10]{amsmath}
\usepackage{amsfonts}
\usepackage{times}
\usepackage[usenames]{color}
\usepackage{verbatim}
\usepackage{array}
\usepackage{cite}
\usepackage{ellipsis}
\usepackage{authblk}

\begin{document}

\title{CellTV - on the Benefit of TV Distribution over Cellular Networks: A Case Study}

\author{Lei~Shi,%
        ~Evanny~Obregon,
        ~Ki~Won~Sung,~\IEEEmembership{Member,~IEEE,}%
        ~Jens~Zander,~\IEEEmembership{Member,~IEEE,}%
        ~Jan~Bostrom
\thanks{Manuscript received February 11, 2013; revised November 4, 2013; accepted November 11, 2013. This work was partly supported by Swedish Post and Telecom Authority (PTS) and Wireless@kth.}%
\thanks{L. Shi, E. Obregon, K. W. Sung, and J. Zander are with Wireless@kth, KTH Royal Institute of Technology, Stockholm, Sweden (e-mail: lshi@kth.se, ecog@kth.se, sungkw@kth.se, jenz@kth.se). J. Bostrom is with PTS, Stockholm, Sweden (e-mail:jan.bostrom@pts.se). The corresponding author is K. W. Sung (e-mail:sungkw@kth.se).}%
\thanks{Color versions of one or more of the figures in this paper are available online at http://ieeexplore.ieee.org.}%
\thanks{Digital Object Identifier xx.xxxx/TBC.2013.xxxxxxx}%
}

\markboth{IEEE Transactions on Broadcasting, Vol.~x, No.~x, xxx~2013}%
{Shi \MakeLowercase{\textit{et al.}}: CellTV - on the Benefit of TV Distribution over Cellular Networks: A Case Study}

\maketitle
\begin{abstract}
As mobile IP-access is becoming the dominant technology for providing wireless services, the demand for more spectrum for this type of access is increasing rapidly. Since IP-access can be used for all types of services, instead of a plethora of dedicated, single-service systems, there is a significant potential to make spectrum use more efficient. In this paper, the feasibility and potential benefit of replacing the current terrestrial UHF TV broadcasting system with a mobile, cellular data (IP-) network is analyzed. In the cellular network, TV content would be provided as {one} of the services, here referred to as \textit{CellTV}. In the investigation we consider typical Swedish rural and urban environments. We use different models for TV viewing patterns and cellular technologies as expected in the year 2020. Results of the quantitative analysis indicate that CellTV distribution can be beneficial if the TV consumption trend goes towards more specialized programming, more local contents, and more on-demand requests. Mobile cellular systems, with their flexible unicast capabilities, will be an ideal platform to provide these services. However, the results also demonstrate that CellTV is not a spectrum-efficient replacement for terrestrial TV broadcasting with current viewing patterns (i.e. a moderate number of channels with each a high numbers of viewers). In this case, it is doubtful whether the expected spectrum savings can motivate the necessary investments in upgrading cellular sites and developing advanced TV receiver required for the success of CellTV distribution.
\end{abstract}

\IEEEpeerreviewmaketitle

\section{Introduction}
\label{sec:intro}
\subsection{Background}
\IEEEPARstart{E}{fficient} use of radio spectrum is considered an essential ingredient of future mobile broadband (MBB) provisioning with exploding capacity demand, in particular for IP-based mobile data. Since IP-access is not tied to a single service, building a single access network, instead of the current plethora of dedicated, single-service systems ("one-trick ponies"), provides economies of scale when it comes to infrastructure deployment as well as a significant potential to make more efficient use of the spectrum.

The UHF broadcasting band is one of the spectrum bands that have attracted special attention due to its favorable  propagation characteristics. In response to the increasing importance of mobile service and its demand for high quality spectrum in sub-1GHz band, the spectrum band between 790 and 864 MHz have been reallocated from TV broadcasting to MBB in Europe by 2013. Despite the loss of 100 MHz spectrum, the digital switchover to DVB-T has revitalized digital terrestrial TV (DTT) broadcasting industry. By March 2013, DTT broadcasting is used in 40\%  of the households in Europe for receiving TV, establishing itself as the most popular platform for TV reception (compared to satellite TV 23\%, Cable TV 19\%) \cite{TNS}. However, DTT take-up varies significantly across Europe, from rather marginal figure in Germany \cite{Reimers}, to over 90\% in Spain \cite{March}.

Meanwhile, the consumption pattern for audio-visual services is shifting rapidly. High definition (HD) and 3D contents are getting increasingly popular, but more importantly the demand trends seem to be shifting towards more diversified contents. Video on-demand (VoD) service has begun to challenge the dominance of linear broadcasting. In fact, audio-visual and Internet data services are increasingly consumed in a unified way. 
It is a challenging issue for the DTT broadcasting industry to meet the growing trends towards 'long-tail' VoD and to face the competition from IPTV, cable TV and satellite TV.

On the other hand, the MBB industry has experienced explosive growth in the last decades. The data traffic is expected to increase by 30 times in five years \cite{Cisco13}, with mobile video constituting two-thirds of the total traffic. 
The increasing amount of high quality audio-visual content accessible via Internet exerts a great pressure on mobile network operators (MNOs) to provide sufficient capacity for multimedia content streaming.

In light of the converging trends of audio-video consumption in both MBB and TV services, World Radio Conference 2012 (WRC-12) allocated the 700MHz band for mobile services on a co-primary basis with DTT broadcasting. This decision has made future authorization for mobile use in this band easier and more attractive but also casted a great uncertainty to the prospect of DTT service \cite{OFCOM13}. 
The European Commission (EC) has also expressed concerns that 'an early and isolated decision on co-allocation of 700 MHz band as of 2015 in the EU could potentially detract from the more comprehensive and coherent inventory process' which is essential for achieving Digital Agenda Europe' target for ubiquitous broadband coverage with high capacity \cite{RSPG}. Therefore, one of the agenda item in WRC-15 is to discuss the possibility of creating a harmonized spectrum band for a converged all-IP platform for delivering both mobile data and audio-visual service through a progressive re-farming of the UHF broadcasting band.

\subsection{Related work}
Numerous studies have focused on investigating solutions for enhancing the utilization of the UHF broadcasting band.
One idea is using the so-called 'TV White Space' on a secondary basis without affecting the normal DTT broadcasting service \cite{maz}. Early in 2010, the Federal Communications Commission (FCC) in the USA announced permission for unlicensed secondary devices assisted by Geo-location database to operate in the TV band \cite{fcc}, while the European regulators have developed their own frameworks for regulating the secondary access \cite{ofcom}\cite{ecc}. Although these pioneering efforts led by the regulators have created high expectations for the secondary access in TV bands \cite{sahai}, quantitative analysis from recent studies has discovered that TV White Space is not suitable for secondary system providing wide-area coverage due to the interference constraint to primary TV receivers \cite{EAB_DYSPAN12} \cite{Lei_DYSPAN12}\cite{jens13}. Only short range systems with smaller interference footprint can efficiently exploit the local secondary spectrum opportunity \cite{Lei_DYSPAN12}. Besides, a possible reallocation of 700 MHz band would greatly affect the amount of available 'TV White Space'.

Another, more radical, approach currently discussed is to re-purpose the UHF band for a cellular, IP-based system to distribute TV contents over this infrastructure as one of many services, and thereby effectively replace the traditional DTT network. Our expectation is that delivering TV service over cellular networks will require less spectrum than current DTT network for the same service offering and quality. At the same time this solution is more flexible and will allow other services to be provided in parallel. One of the enablers is the Evolved Multimedia Broadcast/Multicast Service (eMBMS) introduced in 3GPP LTE (Long Term Evolution) radio technology for point-to-multipoint or multipoint-to-multipoint service over a single frequency network (SFN)\cite{3GPP}. Through tight time synchronization, the TV contents can be broadcasted over a SFN with high spectrum efficiency. Furthermore, additional features such as localized contents distribution and on-demand services are made possible by adopting the cellular infrastructure, and thus considerably improve the flexibility of the TV service. However, as the cost of implementing such system can be considerably high, it would be difficult to motivate the investment unless significant benefit is foreseen. 

The idea of distributing TV contents using a cellular structure was first mentioned in \cite{lery} for coverage extension using relays. Recent studies have mainly focused on analyzing requirements and capacity limits for delivering mobile TV over an OFDMA-based cellular network. In \cite{Hartung}, the authors  present a system architecture of MBMS in 3G networks, and outline the relevance of applying mixed broadcast/unicast solution when there is a "long tail" of channels requested by few users. Detailed traffic analysis for delivering mobile TV over a hybrid broadcast-unicast deployment have been investigated in \cite{Caterin} and \cite{Lohmar}. The implementation and cost aspect of providing mobile TV service in 3G networks are discussed in \cite{Aurelian_IEEENET}. The convergence of mobile TV service and MBB network in 4G networks is presented in \cite{Jorg_conf11}. In \cite{Rong11}, the authors have developed a general roadmap and analytical models for assessing the network performance in terms of coverage and throughput for different deployment options using advanced features introduced in LTE network.

However, DTT service has a completely different service demand than mobile TV, and also has a significantly higher quality of service requirement. Current DVB-T system offers HDTV program that requires a data rate over 7 Mbps, whereas the data rate of a typical mobile TV transmission is in the range of hundreds of Kbps. Furthermore, the strict coverage requirement of DTT poses a formidable challenge for any attempt to replace it with mobile networks. The same high quality TV programs are supposed to reach the fixed receivers even at the edge of the coverage. On the other hand, fixed TV receivers can rely on more advanced antenna configurations with considerably better performance than mobile receivers. Consequently, the existing results on mobile TV cannot be directly applied to the study on distributing terrestrial TV service over mobile network.

Using LTE technology to provide over-the-air TV service has been proposed in \cite{Reimers} as a 'tower-overlay' system, where the DTT network employs a modified LTE standard for broadcasting TV content to both mobile and fixed reception. Recent studies have considered using not only LTE technology but also cellular infrastructure for providing TV services. In \cite{Huschke}, the amount of spectrum needed for delivering today's over-the-air TV service is calculated by taking different cities in the USA as reference. Its focus is limited to densely populated (urban) areas where typical inter-site-distance (ISD) of cellular networks is smaller than 2km, which ensures good performance of the eMBMS network. Larger ISD which is typical in rural areas would considerably degrade the spectral efficiency of the SFN due to the long propagation delay as shown in \cite{Hartung}, thus requiring far larger amount of spectrum to provide the same service. Therefore, it is not evident that replacing DTT service with mobile networks is feasible based on the results from urban scenarios alone. Besides, the possibility of employing unicast for less popular TV channels is not exploited in this analysis, although it may reduce the spectrum requirement as indicated by results from earlier studies.

\subsection{Contribution}

\begin{figure}[t]
  \centering
  \includegraphics[width=0.48\textwidth]{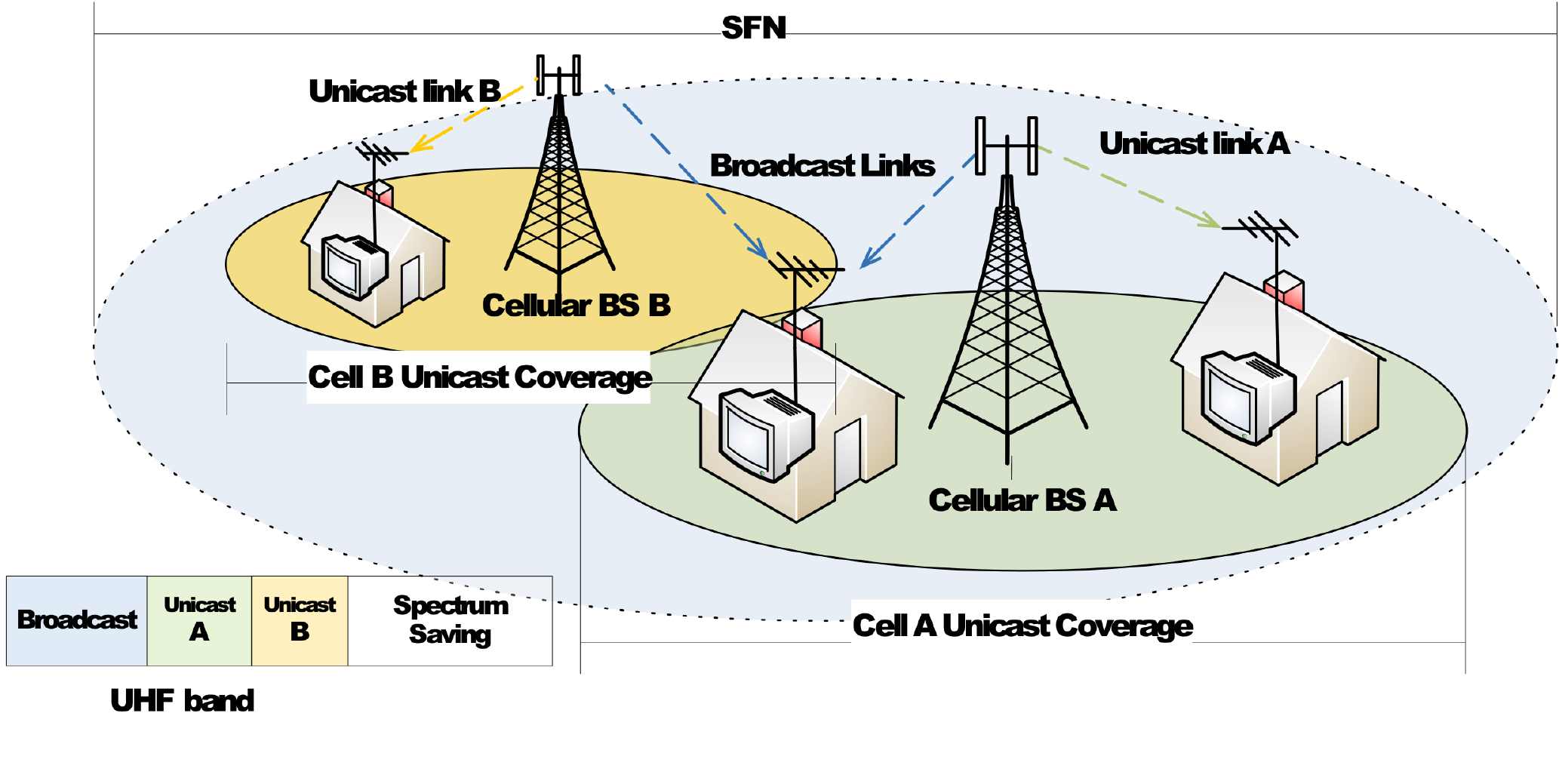}\\
  \caption{Illustration of CellTV system.}\label{fig:CellTV}
\end{figure}

In this paper, we aim to provide a more comprehensive assessment of the potential benefit of CellTV, which is defined as using the cellular infrastructure and technology to deliver terrestrial TV service to fixed receptions as illustrated in Fig. \ref{fig:CellTV}. Two possible architectures of the CellTV concept are investigated. One option is to deliver all TV programs over (several) SFN(s) formed by multiple cellular sites, while the other is to broadcast only the most popular TV programs and distribute the rest of programs via unicast links. The system performance is evaluated in terms of spectrum saving, referring to the portion of spectrum out of 470-790 MHz band that can be vacated for broadband usage. To properly reflect the various spectrum demand of unicast viewers in different situations, multi-Erlang model is applied to analyze the capacity and the spectrum requirement of the hybrid system. The investigation targets the year 2020 with moderate assumptions on the cellular technologies development, such as advanced MIMO (multiple input and multiple output) and enhanced modulation/coding schemes. We based our numerical analysis on the statistics of cellular network deployment and DTT service in Sweden, which has one of the best DTT coverage as well as mobile coverage in Europe. Sweden also consists of a good mixture of sparsely populated rural areas and dense urban cities. Lastly, we study the potential impact of the possible changes in the number of TV channels, the terrestrial TV service penetration, and its consumption pattern in the coming years.

The remainder of this paper is organized as follows: Section \ref{sec:Problem Formulation} defines the objective of the study and describes the expected requirements for over-the-air TV service in 2020 in Sweden. The modeling of CellTV and the calculation of its spectrum requirement are explained in Section \ref{sec:spectrum}. Then, Section \ref{sec:result} describes the representative Swedish scenarios for numerical evaluation and the major results. Finally, the main conclusion and implications are discussed in Section \ref{sec:concl}.

\section{Problem Formulation}
\label{sec:Problem Formulation}

The aim of this study is to quantify the required spectrum for replacing DTT network with distributing TV service using the cellular infrastructure and technology. The potential benefit of CellTV is evaluated by comparing its required spectrum to the amount of spectrum currently allocated for DTT networks.

\subsection{Analysis Scenarios}
Sweden is chosen for our case study for its diverse morphologies types as well as its good coverage of both DTT system and mobile network. We focus the investigation on typical Swedish rural and urban environments, as they represent two distinct cases with regard to the spectrum demand.
\subsubsection{\textbf{Rural}}
 The population density is very low in typical Swedish rural areas, where most of the rural inhabitants rely on over-the-air TV reception. It is estimated that around 60\% of the households in rural areas have subscriptions to the DTT service\cite{Tera}. These households are assumed to receive the TV signal through high gain rooftop antennas, which can be legacy type antennas or advanced multi-antenna units with MIMO capability that are expected to be commercially available by 2020. However, as further cell site acquisition is not likely to happen in rural areas, the limited cellular infrastructure may pose a significant challenge for providing the coverage with intended data rate requirement.

\subsubsection{\textbf{Urban}}
The urban area in Sweden, e.g., Stockholm, has a much higher population density and also a denser cellular base station deployment. As most families in the city have cable connections, the DTT service penetration is estimated to be only 15\%\cite{PTS1}. However, the density of DTT viewers in the city is still considerably higher than the rural area. Rooftop antenna is seldom used in apartment buildings. Instead, we assume that indoor gateways with multiple low gain antennas are used in the urban environment.

\subsection{Requirement for Terrestrial TV Service} \label{sec:TV_model}
\subsubsection{\textbf{Service Availability Requirement}}
In Sweden, the DTT network currently covers more than $99.8\%$ of the inhabited area. It is required that the service availability must be higher than $95\%$ at the TV coverage boundary, which is approximately equivalent to a service availability of $99\%$ within the whole TV coverage area.
The temporal availability is not explicitly defined in the DTT system because the broadcast service is expected to be constantly available. However, with the introduction of unicast for TV distribution, there is a risk of temporary blocking due to fluctuations in the traffic load. Therefore, we assume that a strict requirement on temporal availability, e.g. $99.9\%$, should be imposed on the CellTV system in addition to the coverage requirement.
\subsubsection{\textbf{Terrestrial TV Service in 2020}}

The number of TV programs being simultaneously broadcasted over the DTT network in Sweden is expected to increase slightly in 2020, reaching 60 in total, out of which 36 would be high-definition (HD) programs and 24 standard-definition (SD) programs. For public service and commercial interest, some TV programs may have regional content that differs in each region. We assume that the division of region remains the same as that of today, i.e., at most three intersecting regions at any location within Sweden. 
During the peak hour (8-9 pm), over $40\%$ of the households in Sweden would be watching TV and half of them would tune to the three most popular TV programs. The dimensioning of the CellTV system is based on the peak hour traffic assumption \cite{PTS1}.

\subsection{CellTV Distribution Methods}
In this study, we investigate the feasibility of the CellTV concept defined as using cellular infrastructure to deliver traditional DTT services. Since CellTV is envisaged as a replacement of DTT network, the priority is given to ensuring the service quality for fixed receptions, which corresponds to the primary segments of the DTT users and have stricter requirements on the coverage and quality of service than mobile receptions. We consider two possible operation modes for CellTV: broadcast-only or a mixture of broadcast and unicast.

We envisage a scenario with a shared network among multiple operators to avoid the spectrum wasting situation where each individual MNO would broadcast the same content to its own customer only. There are also other possible service scenarios as discussed in \cite{qualcomm}, ranging from utilizing current LTE standard with carrier aggregation to provide broadcast content to multiple customer bases, to an independent broadcast network operator offering service for different MNOs. Therefore, we assume that, with the business model development, the same content would be broadcasted only once to the customers of multiple MNOs.

\subsubsection{\textbf{Broadcast-Only}}
In this configuration, all TV programs except those with regional content are broadcasted over a large scale SFN formed by a group of cellular base stations transmitting on the same frequencies using eMBMS technology. The TV programs with regional content, on the other hand, are distributed through regional SFNs each operating on a unique set of frequencies.
\subsubsection{\textbf{Hybrid of Broadcast-Unicast}}
This hybrid of broadcast-unicast distribution allows the CellTV system to broadcast only the few popular TV programs over SFNs using eMBMS and deliver the rest of the TV programs as typical video streaming on unicast links. In addition to the streaming of linear TV programs, the cellular unicast also enables enhanced features, such as VoD service.

\subsection{Performance Metric}
The key performance metric is the amount of required spectrum, $BW_{req}$, defined as the total amount of radio spectrum to be allocated for the CellTV system in order to provide the same level of service offered by DTT networks throughout Sweden. The frequency band in question is the UHF band between 470 MHz and 790 MHz, which is assumed to be no longer occupied by DTT networks.

Spectrum saving is simply defined as the difference between the amount of spectrum allocated to DTT system and the amount of spectrum required by the CellTV system.
\begin{equation}\label{BW_save}
  	BW_{save}=320-BW_{req}	\mathrm{(MHz)}
\end{equation}	
A positive value of spectrum saving indicates the potential gain of the CellTV distribution, while a negative one may imply the infeasibility of providing the CellTV service within the UHF broadcasting band.

\subsection{Evaluation Methodology}
\label{sec:Methodology}

The quantitative analysis performed in this study can be divided into two phases:
\begin{enumerate}
  \item 	{Selection of Representative Cases}: first, specific locations that are deemed as the most problematic for cellular TV distribution are selected from Swedish rural and urban areas, respectively. Then, representative parameters are extracted from the base station deployment and demographics data of the selected areas.

  \item {Calculation of Spectrum Requirement}: based on these representative parameters, the evaluation scenario is constructed with a regular deployment of cellular sites and uniformly distributed TV receivers. Then, the required spectrum for the CellTV system for the particular setting is calculated using the analytical tools and simulation models described in Section \ref{sec:spectrum}.
\end{enumerate}

\section{Required Spectrum for CellTV Distribution }\label{sec:spectrum}
\subsection{Broadcast-only CellTV Distribution}
\subsubsection{\textbf{Spectrum Allocation for Broadcast-only}}\label{sec:BW_broad}

Multicast-broadcast over a single frequency network (MBSFN) introduced in eMBMS enables multiple transmissions from multiple base stations over the same frequency channel, which is seen from a receiver as a single transmission subject to a severe multi-path propagation. Tight time synchronization of all base stations is required to overcome the effects of ISI (inter-symbol interference). Due to the long data symbol duration of OFDM, LTE MBSFN considerably mitigates ISI effects when the delay spread is relatively small. The propagation delay of the transmitted signals has a critical impact on the performance of MBSFN. As opposed to traditional unicast transmission, portion of the transmitted signals that arrives within a certain duration is considered constructive interference or gain \cite{Rong08}.

\begin{figure}[t]
  \centering
  \includegraphics[width=0.48\textwidth]{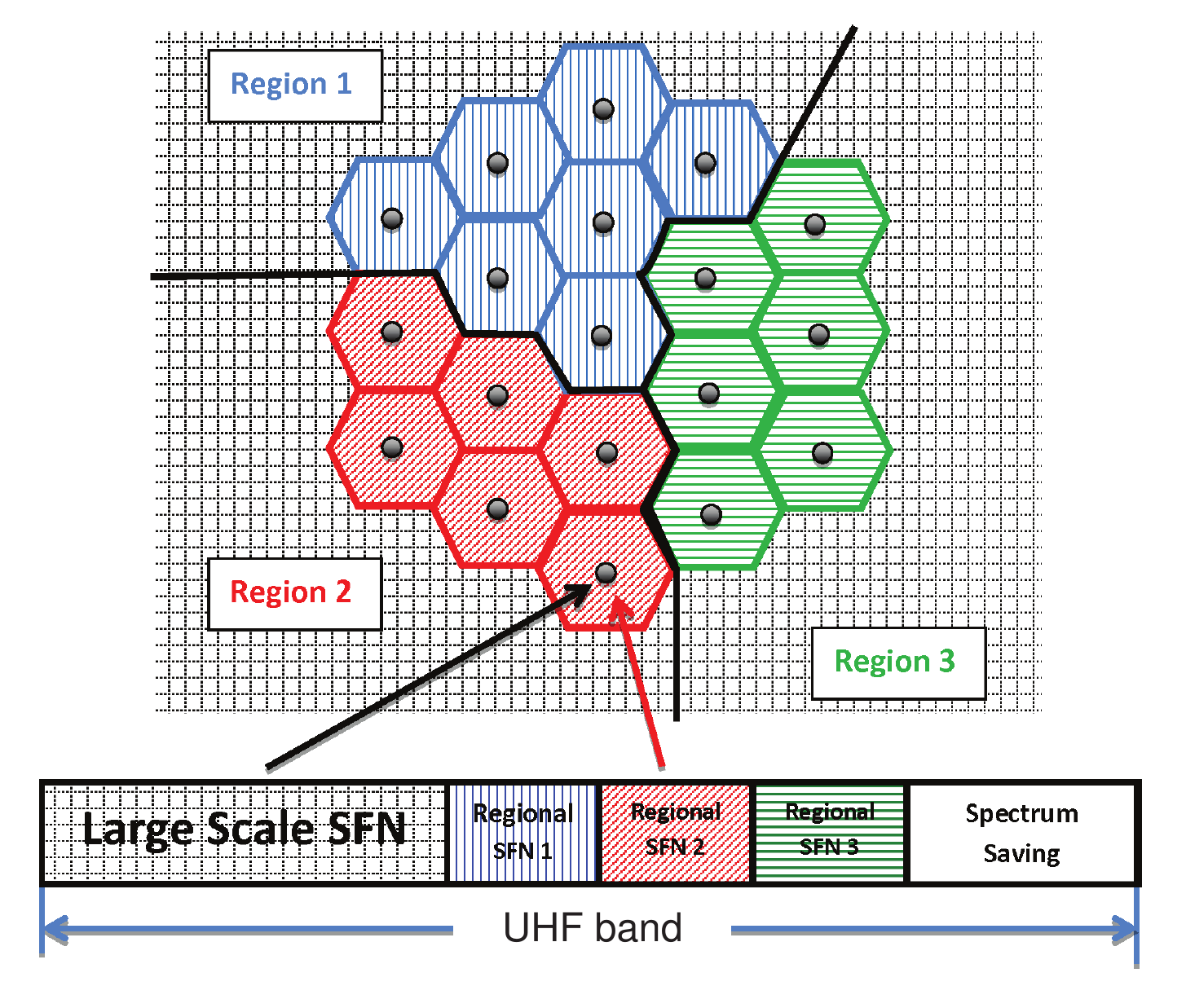}\\
  \caption{Spectrum allocation of CellTV with pure broadcast.}\label{fig:broadcast_diag}
\end{figure}

While the nationwide TV programs can be transmitted over a large scale SFN, the TV programs with regional content must be transmitted separately over different SFNs operating on different set of frequency channels in each geographical region. Fig. \ref{fig:broadcast_diag} illustrates the spectrum allocation for different SFNs in CellTV network.

Spectral efficiency of regional channels at the border of a regional SFN is lower than that of the channels belonging to the large scale SFN because of less SFN gain. Therefore, the regional border cell has the highest spectrum requirement, representing the worst-case scenario. Denoting $ESE_{min}^R$ and $ESE_{min}^L$ as the effective spectral efficiency (ESE) for the SFN link at the cell border for regional channels and national channels, the total bandwidth required for a border cell can be calculated by,	
\begin{equation}\label{BW_broad}
  BW_{req}^B=\frac{\eta_{HD}^L R_{HD}+\eta_{SD}^L R_{SD}}{ESE_{broad,min}^L}+X\frac{\eta_{HD}^R R_{HD}+\eta_{SD}^R R_{SD}}{ESE_{broad,min}^R },
\end{equation}
where $\eta^L$ and $\eta^R$ denote the numbers of TV programs distributed in large scale SFN and regional SFNs, respectively. The subscripts HD and SD are used to distinguish between HD and SD TV programs. $R_{SD}$ and $R_{HD}$ are the rate requirements for SD and HD TV programs, respectively. $X$ is the number of intersecting regions around the studied area.

\subsubsection{\textbf{SINR for Broadcast over SFN}}
Assume that the target user is located in cell $0$ at a distance $r_0$ from base station $0$ and at distance $r_i$ from an arbitrary base station $i\neq0$ in cell $i$. The constructive portion of a received SFN signal depends on the propagation delay $\tau=(r_i-r_o)/c$, where $c$ is the speed of light. For a given $\tau$, the weight function of the constructive portion of a received SFN signal is \cite{Mogensen}\cite{jens}:
\begin{equation}\label{eqn:weight_function}
 \omega(\tau)=\left\{
                \begin{array}{ll}
                    0, & \hbox{$\tau<-T_{u}$;} \\
                  1+\frac{\tau}{T_u}, & \hbox{$-T_{u}\leq\tau<0$;} \\
                   1, & \hbox{$0\leq\tau<T_{CP}$;} \\
                  \frac{1-(\tau-T_{CP})}{T_u}, & \hbox{$T_{CP}\leq\tau<T_{CP}+T_u$ ;} \\
                  0, & \hbox{otherwise,}
                \end{array}
              \right.
\end{equation}
where $T_u$ is the length of the useful signal frame and $T_{CP}$ is the length of the cyclic prefix. Due to multipath propagation, multiple copies of a signal could arrive to the receiver. Then, the weighted function should be calculated for each multipath signal. Typically OFDMA attenuates the impact of fast fading by guaranteeing that all multipath signals arrive within the cyclic prefix \cite{Rong11}. The raw SINR of a user in cell $0$ is given by:
\begin{equation}\label{eqn:SINR_MBSFN}
  SINR_{broad}=\frac{\sum_{i=0}^m\frac{\omega(\tau_i )\bar{P}}{q_i} }{\sum_{i=1}^m\frac{(1-\omega(\tau_i))\bar{P}}{q_i} +N_0},
\end{equation}
where $\tau_i$ is the propagation delay, $\bar{P}$ is average power associated with base station $i$, and $q_i$ represents the propagation loss to the base station $i$ which accounts for distance-based path loss and shadowing. The total number of cells in the MBSFN area is given by $m$.
It should be noted that this SINR calculation may lead to optimistic results as pointed out in \cite{Plets10} \cite{Plets11}. The performance of SFN in a realistic scenario would be less homogeneous and affected by non-ideal receiver response function. To compensate any potential overestimation of SINR performance, we have made conservative assumptions on the loss factor in spectral efficiency calculation in (5).

\subsubsection{	\textbf{Effective Spectral Efficiency}}
\begin{table*}[t]
\caption{Bandwidth efficiency for MBSFN and unicast \cite{Adachi,Zheng}.}\label{table:eff}
\centering
\begin{tabular}{|c|c c|c c|}
  \hline
    & Rural broadcast & Urban broadcast & Rural unicast & Urban unicast \\
    \hline
  ACLR overhead & 0.1 & 	0.1 & 0.1 & 	0.1 \\
 Cyclic prefix overhead&	0.2 &	0.07&	0.2 &	0.07\\
Pilot and control overhead&	0.1	&0.1&	0.3 &	0.3\\
$\beta_{eff}$ &	0.65&	0.75 & $0.5\cdot \min(M_T,M_R)$	& $0.59\cdot \min(M_T,M_R)$\\
  $\xi_{eff}$ & $M_TM_R/2$ & $M_TM_R/2$&	0.5&	0.5\\
\hline
\end{tabular}
\end{table*}
In order to calculate the Effective Spectral Efficiency (ESE), we adopt a simplified model based on the Shannon formula. To draw a realistic link performance of future cellular system in relation to the Shannon capacity bound, we employ two parameters: bandwidth efficiency ($\beta_{eff}$) and SINR implementation efficiency ($\xi_{eff}$) \cite{Mogensen}. Then, the modified Shannon capacity formula is expressed as follows:
\begin{equation}\label{ESE}
  	ESE_{broad}(\textrm{bps/Hz})=\beta_{eff}\log_2⁡[(1+\xi_{eff} \overline{SINR}_{broad})].	
\end{equation}
Here, $\overline{SINR}_{broad}$ is computed when the wireless link is in deep fading, which represents 5dB loss in the raw SINR (${SINR}_{broad}$). This assumption is made to account for the impact of fast fading \cite{Adachi}. The parameter $\beta_{eff}$ is determined by the adjacent channel leakages ratio (ACLR) requirements and protocol overheads. $\xi_{eff}$ corresponds to the SINR which is mainly affected by the modulation, coding, and MIMO modes. The maximum modulation order is assumed to be 512QAM (a maximum spectral efficiency of 9 bps/Hz per stream). Table 1 shows the parameters and values used for calculating $\beta_{eff}$ for rural and urban scenarios. Notice that for the broadcast-only case, we have neglected the gains related to dynamic beamforming due to the lack of user feedback. Therefore, we consider that $\xi_{eff}$ for broadcast systems mainly depends on the spatial and polarization diversity gain, which is proportional to the number of transmitting and receiving antennas, $M_T$ and $M_R$, respectively.

In order to provide the same coverage quality of the traditional DTT service, the CellTV network must cover all the inhabited area (households with permanent addresses) in Sweden with $99\%$ reception probability. In other words, the bandwidth allocated for a given TV program should be sufficient to achieve the required data rate even for a user experiencing the lowest 1 percentile SINR.

\subsection{Hybrid Broadcast-Unicast CellTV Distribution	}
\subsubsection{\textbf{Spectrum Allocation for Hybrid Broadcast-Unicast}}
\begin{figure}[t]
  \centering
  \includegraphics[width=0.48\textwidth]{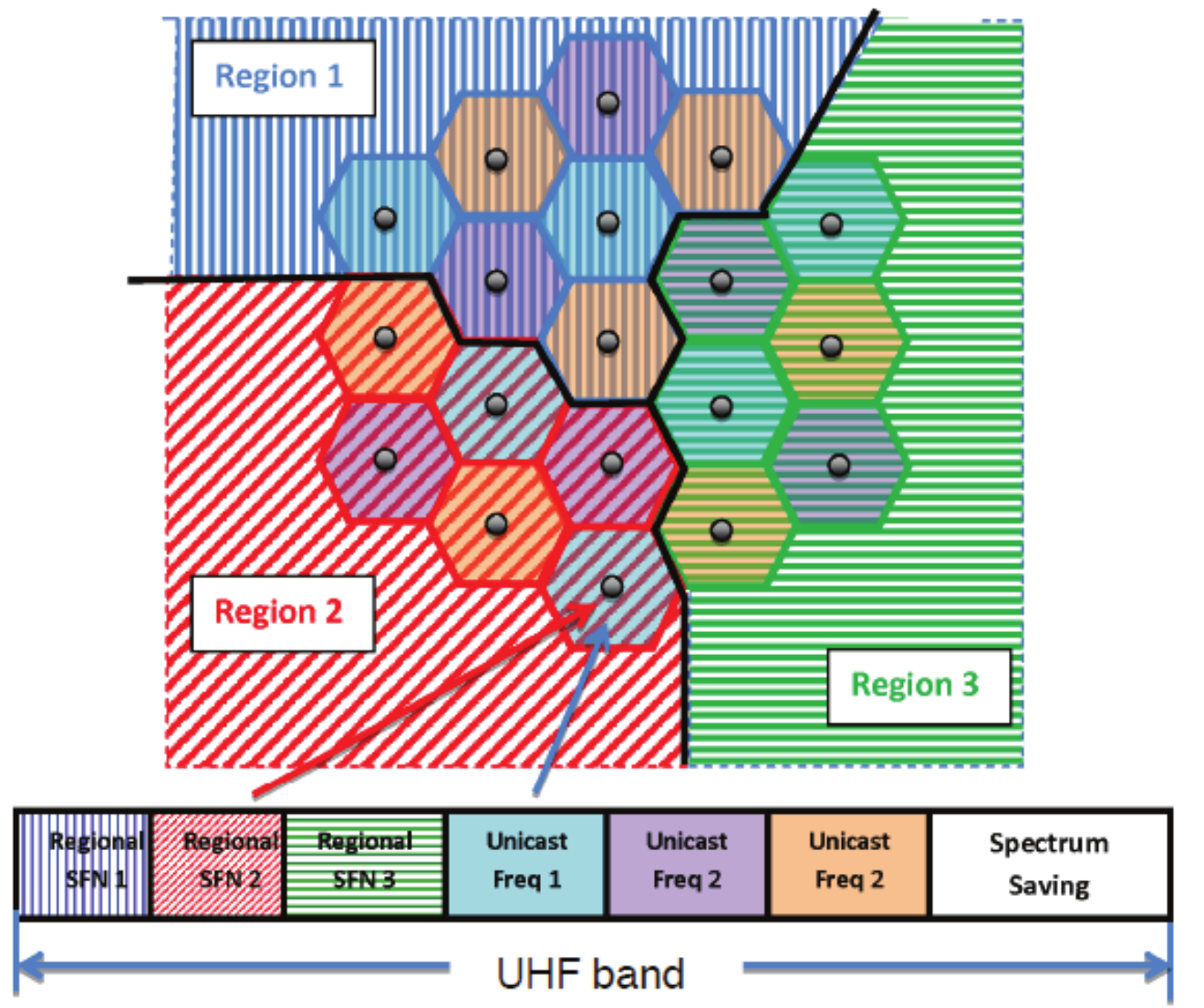}\\
  \caption{Spectrum allocation for CellTV with hybrid broadcast-unicast operation.}\label{fig:unicast_diag}
\end{figure}
In this hybrid distribution mode, the TV programs with the largest amount of viewers are broadcasted over SFNs. For the worst-case, these TV programs are assumed to contain regional contents. On the other hand, less popular TV programs are delivered via unicast links as typical point-to-point video streaming. For these transmissions, frequency reuse of $K$ could be applied to limit the co-channel interference from other cells. Fig. \ref{fig:unicast_diag} illustrates the spectrum allocation in the hybrid broadcast-unicast system for the case of spectrum reuse three.

Assuming per-cell spectrum requirement for unicast to achieve sufficiently low blocking probability is $BW_{uni}$, the total bandwidth required for the hybrid distribution is given by
\begin{equation}\label{eqn:BW_uni}
  	BW^H_{req}=X\cdot BW_{broad}+K\cdot BW_{uni},	
\end{equation}
where $BW_{broad}$ is bandwidth for broadcasted channels in three regional SFNs derived by using the methodology in section \ref{sec:BW_broad}. In our study, $K=3$ is adopted.

\subsubsection{\textbf{Traffic Model for TV Viewing}}
As opposed to the broadcast case, the bandwidth required for unicast is dependent on the number of TV viewers per cell. Assume that the number of active TV viewers in a cell follows a Poisson distribution, $N\in Poi(t_s,\lambda)$, with $t_s$ being the average session length and $\lambda$ the arrival rate. Based on statistics of the average ISD and population density \cite{PTS2}\cite{PTS3}, the average number of active TV viewers in a cell $E\{N\}$ can be estimated by the product of the average TV viewing ratio and the number of terrestrial TV receivers within the cell coverage. Then, the arrival rate of the TV viewers is given by
\begin{equation}\label{lambda}
 	\lambda=E\{N\}/t_s.	
\end{equation}

Let $\Omega$ denote the set of TV programs delivered by CellTV (both via broadcast and unicast). An active TV viewer may select the $i^{th}$ TV program with probability $P_i, i\in\Omega$. This selection probability can be approximated by the popularity of that TV program and $\sum_{i\in\Omega}P_i =1$. Within each session, the TV viewer can switch between different TV programs (either delivered by broadcast or unicast link) and spend $t_c$ on average for each sub-session. At the end of each sub-session, the viewer may turn off the TV with probability $P_e$. The total session time for such system can be modeled by Coxian distribution. It follows that $P_e=t_c/t_s$. As proved in \cite{Caterin}, the stationary distribution of the numbers of active unicast viewers is given by
\begin{equation}\label{eqn:N_pop}
  	N_{uni}\sim Poi(\rho_{uni}),~\rho_{uni}=t_s\lambda\sum_{i\in\Omega_{uni}}P_i.
\end{equation}
Here, $\rho$ is the traffic intensity. $\Omega_{uni}\subseteq\Omega$ is the set of unicast TV programs. Note that the numbers of viewers watching different sets of TV programs are independent of each other, and are not affected by the sub-session duration \cite{Caterin}.

Assume the bandwidth required for the $i^{\textrm{th}}$ unicast link is $b_i$. If any viewer watching unicast program fails to secure $b_i$, this cell is considered in blocking state. The blocked viewers keep attempting to enter the desired channels until either they success or their sessions time out. The blocking probability is
\begin{equation}\label{eqn:block}
  P_{block}=\Pr(\sum_{i=1}^{N_{uni}}b_i >BW_{uni}) ,
\end{equation}
where $BW_{uni}$ satisfy $P_{block}\leq 0.1\%$. This value is chosen to represent the strict requirement on the blocking of unicast TV service, although our analysis also shows that the $BW_{uni}$ is not very sensitive to the blocking requirement.

\subsubsection{\textbf{Multi-Erlang Analysis}}
The bandwidth to be allocated to the unicast service can be obtained by testing different values in Monte Carlo simulations until the blocking requirement defined in (\ref{eqn:block}) is satisfied. However, such iterative process would require extensive simulations. To reduce the computation complexity, we approach the problem with multi-Erlang analysis to solve it analytically.

To construct a multi-Erlang system, the unicast viewers in a cell are divided into $K$ different streaming classes according to their required bandwidths, such that $b_k\geq b_i>b_{k+1}$ ($k=1,2,...K$). The traffic intensity corresponding to each class is given by $\rho_k$. Since $b_i$ is directly dependent on the link SINR, the streaming classes can be defined according to the SINR distribution in the cell. Assuming that the TV viewers are uniformly located inside the TV coverage, subject to uncorrelated shadow fading, the streaming class corresponding to $(k-1)\Delta\%\sim k\Delta\%$ of the SINR distribution has the traffic intensity defined as
\begin{equation}\label{rho}
   \rho_k=\frac{\Delta}{100}t_s\lambda\sum_{i\in\Omega_{uni}}P_i.
\end{equation}
Note that all users with SINR lower than the minimum requirement are considered in outage and do not contribute to the traffic in the system.
To further distinguish the different rate requirements of HD and SD programs, each class can be divided into two subclass as
\begin{equation}\label{hd/sd}
  \rho_k^{HD/SD}=\rho_k\frac{\sum_{i\in\Omega_{uni}^{HD/SD}}P_i}{\sum_{i\in\Omega_{uni}}P_i}.
\end{equation}
The blocking probability, which is equivalent to the portion of time in blocking state, is thus given by
\begin{equation}\label{eqn:block2}
  P_{block}=\Pr\{\sum_{k=1}^K\ N_k b_k>BW_{uni}\},
\end{equation}
where $N_k$ is the number of active user in class $k$ ($N_k~Poi(\rho_k)$).

The blocking probability can be obtained by following Kaufman-Roberts recursion \cite{Roberts} briefly illustrated below:
\begin{itemize}
  \item Find a small unit value $\delta$ such that $b_k\approx b_k'\delta$, $k=1,2,...,K$ and $BW_{uni}\approx C\delta$ with both $b_k'$ and $C$ being integers.
  \item Define $G(c)$ following the recursion process given by
      \begin{equation}\label{G}
     G(c)=\frac{1}{c}\sum_{k=1}^K\rho_kb_k'G(c-b_k'),
\end{equation}
which is initialized by $G(0)=1$, and $G(c)=0$ when $c<0$.
  \item Solve $G(c)$ for $c=1,2...C$.
  \item Obtain the blocking probability as
  \begin{equation}\label{eqn:Block3}
    P_{block}=\sum_{k=1}^K\frac{\sum_{c=C-b_k'+1}^CG(c)}{\sum_{c=0}^CG(c)}.
  \end{equation}
\end{itemize}

\subsubsection{\textbf{Unicast SINR}}
To obtain the unicast link SINR distribution, let us consider an arbitrary viewer at location $r_i$ whose raw SINR can be expressed as
\begin{equation}\label{SINR_uni}
  	SINR_{uni}(r_i,X)=\frac{\bar{P}/q_0 (r_i)}{\sum_{l=1}^{m'}X_l\bar{P}/q_l (r_i)+N_0},
\end{equation}
where $X$ is the interference collision vector conditioned on the network load $x$ and $m'$ is the number of interfering base stations (sites allocated with the same spectrum for unicast). Then, the spectral efficiency is derived using the same model as in the broadcast case:
\begin{equation}\label{ESE_uni}
\begin{aligned}
  	ESE_{uni}(r_i,X)(\mathrm{bits/s/Hz})=\\
  \beta_{eff}\log_2⁡[1+\xi_{eff} \overline{SINR}(r_i,X)].
    \end{aligned}
\end{equation}
Here, $\beta_{eff}$ is modified to reflect the beamforming gain and the increased control overhead. $\xi_{eff}$ is also changed from diversity gain to represent the MIMO implementation loss instead. The parameter settings are summarized in Table \ref{table:eff}.

The network load $x$ in the system is obtained by solving the fixed point equation
\begin{equation}\label{x}
\begin{aligned}
  x=&\min⁡\left[(\rho_{HD} R_{HD}+\rho_{SD} R_{SD})\cdot\right.\\
  &\left.\int_0^R\!\frac{2r\ \mathrm{d}r}{R^2 \sum_X[Pr⁡(X|_x)BW_{uni}ESE(r,X)]},1\right].
  \end{aligned}
\end{equation}
Here, $ESE(r,x)$ is averaged over shadow fading. Note that the network load $x$ is thus depending on the total bandwidth available for unicast $BW_{uni}$ and the traffic intensity of unicast TV viewers in a cell watching either HD or SD programs ($\rho_{HD}$ and $\rho_{SD}$ ).

\section{Numerical Evaluation}	\label{sec:result}						
\subsection{Parameter Settings}			

\begin{table*}
\caption{TV service and consumption.}\label{table:service}
\centering
\begin{tabular}{|c|c|}
  \hline
Parameters&	Values \\
\hline
Peak hour TV consumption &$40\%$ of total population\\
TV per household	&2\\
Population per household	&2.1\\
Number of HD programs	&36 (in 2020)\\
Number of SD programs	&24 (in 2020)\\
Data Rate requirement for one HD program &7.14Mbps	\\
Data Rate requirement for one SD program &1.83Mbps	\\
Number of programs with regional content in broadcast-only	&3 (HD) \\
Number of programs delivered via broadcast in hybrid operation	&3 (HD, accounts for $50\%$ of the viewers)\\
\hline
\end{tabular}
\end{table*}

\begin{table*}
\caption{Simulation parameters for rural scenario.}\label{table:rural}
\centering
\begin{tabular}{|c|c|c|}
  \hline
Parameters&	LTE Outdoor Base Station&	Receiver antenna \\
\hline
Number of antennas	&4, 8	&1, 4, 8\\
Antenna gain	&15dBi	&8dBi \\
Transmit power	&46dBm/20MHz/antenna &N/A	\\
Antenna height	&90m	&10m\\
Tilt (down)	&2.5degrees&N/A \\
Polarization	&+/- 45 cross-polarized	&Horizontal polarization: ITU-R BT.419 \cite{BT419}\\
Noise figure	&N/A	&7dB\\
Noise floor	&N/A &-94dBm/20MHz	\\
\hline
ISD range	&\multicolumn{2}{c|}{4km - 16km} \\
Population density 	&\multicolumn{2}{c|}{$1 \mathrm{~inhabitants} /\text{km}^2$} \\
Terrestrial TV service penetration	&\multicolumn{2}{c|}{$60\%$} \\
Wall attenuation  &\multicolumn{2}{c|}{0dB} \\
  \hline
\end{tabular}
\end{table*}

\begin{table*}
\caption{Simulation parameters for urban scenario.}\label{table:urban}
\centering
\begin{tabular}{|c|c|c|}
  \hline
Parameters&	LTE Outdoor Base Station&	Indoor gateway \\
\hline
Number of antennas	&4, 8	&1, 4, 8\\
Antenna gain	&15dBi	&0dBi \\
Transmit power	&46dBm/20MHz/antenna &N/A	\\
Antenna height	&30m	&1.5m\\
Tilt (down)	&2.5degrees&N/A \\
Polarization	&+/- 45 cross-polarized	&Vertical polarization\\
Noise figure	&N/A	&10dB\\
Noise floor	&N/A &-91dBm/20MHz	\\
\hline
ISD range	&\multicolumn{2}{c|}{100m - 1500m} \\
Population density 	&\multicolumn{2}{c|}{$5000 \mathrm{~inhabitants}/\text{km}^2$} \\
Terrestrial TV service penetration	&\multicolumn{2}{c|}{$15\%$} \\
Wall attenuation  &\multicolumn{2}{c|}{10dB} \\
  \hline
\end{tabular}
\end{table*}

\begin{figure}[t]
  \centering
  \includegraphics[width=0.30\textwidth]{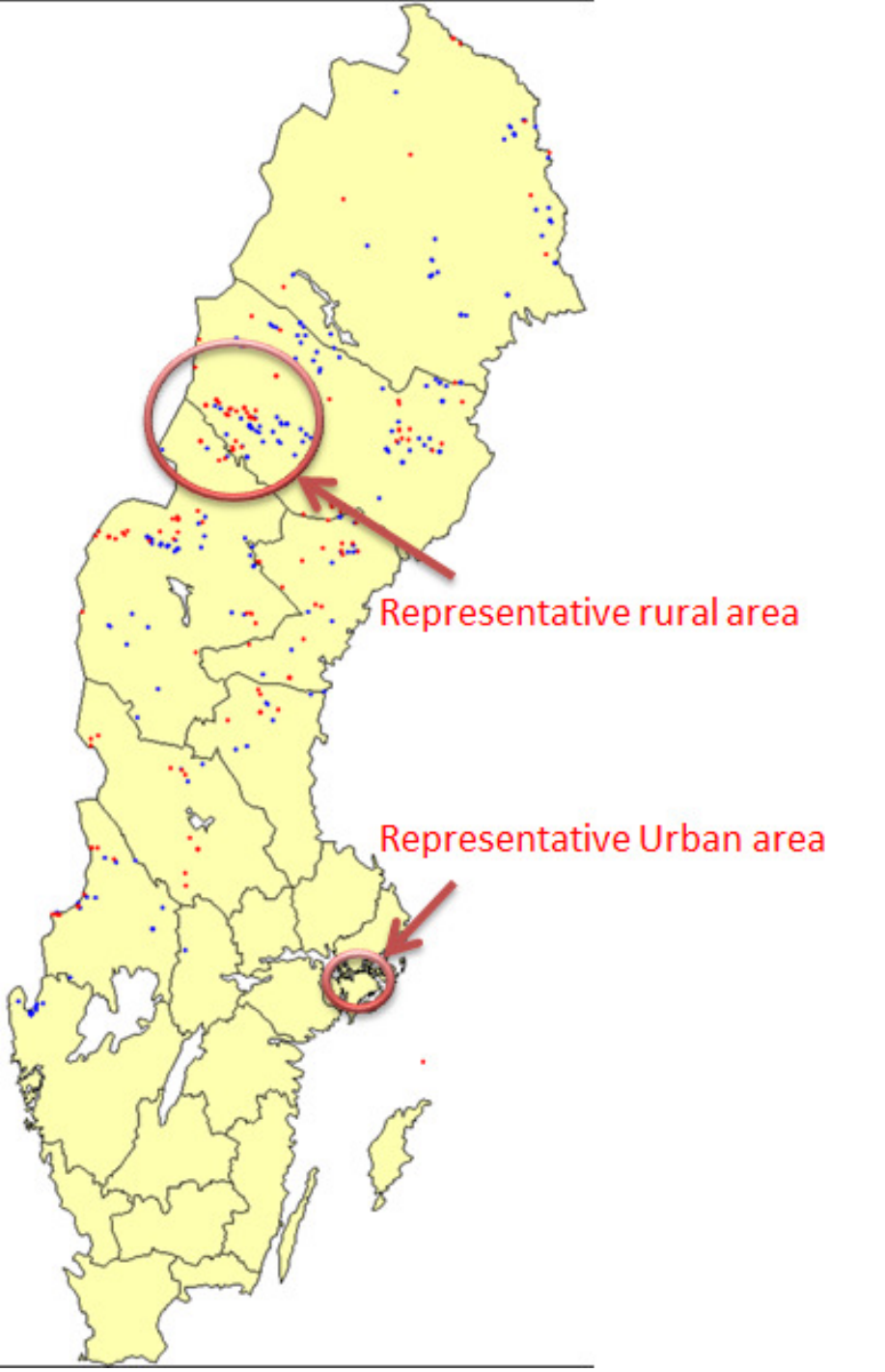}\\
  \caption{Area without broadband connection and selected investigation areas \cite{PTS4}; the blue and red dots on the map indicate the locations where the broadband connection are lacking (blue for 2009 record and red for 2010 record).}\label{fig:Area}
\end{figure}

For numerical evaluation, the simulation scenarios are created based on the typical settings of Swedish rural and urban areas. The choice of Stockholm for the urban scenario is straightforward. Less obvious is the selection of the rural area because Sweden contains vast area with sparse population. The most interesting scenario is identified as the area with the most problematic broadband coverage according to the recent Swedish Post and Telecom Authority (PTS) report \cite{PTS4}. The selected area for study is outlined in red circles in Fig. \ref{fig:Area}.

Having identified the areas of investigation, we extract representative parameters from the base station deployment and the demographics data and construct the simulation environment as a regular hexagonal cellular deployment. The number of TV receivers in a cell is derived from the population density. On average, one Swedish household consists of 2.1 populations and each household possesses two TVs (see Table \ref{table:service}). Since there is a maximum of three intersecting regions throughout Sweden, only three sets of different frequency channels are needed for the regional SFNs ($X=3$). We assume that the bit rate requirement for SD programs is 1.83Mbps, corresponding to the video format of 576i and coding format of H.264/AVC (MPEG4). For HD program, we assume that a minimum bit rate of 7.14Mbps is required, corresponding to 1080i or 720p video format using H.264/AVC (MPEG4), or 1080p video format using high efficiency video coding (HEVC). The simulation parameters for the rural and urban scenarios are summarized in Table \ref{table:rural} and Table \ref{table:urban}, respectively. For the hybrid operation, only the top three TV programs accounting for $50\%$ of viewing ratio are broadcasted by regional SFNs. All other channels are delivered via unicast links in a cellular network with frequency reuse $K=3$.

Notice that we assume the CellTV may utilize the existing GSM sites for delivering the TV service because it provides a more homogenous coverage than the existing UMTS sites. The antenna height of 90 m for the base station is normal in rural Sweden to maximize the reach in these extremely sparse areas. We further assume the cellular network would be equipped with significantly larger backhaul capacity than it is available today to support the growing data traffic including the provisioning of audio-visual content.

On the receiver side, we consider the cases that either all the users are using legacy antenna or they have replaced it with MIMO capable new antennas. In rural areas where there is likely a line-of-sight between transmitter and receiver, a spacing of more than 2-3 meters \cite{MIMO} would be required to avoid significant spatial correlation. Due to the space limitation, it might not be practical to install an antenna with more than four uncorrelated branches on the rooftop or sidewall of a typical household (a four-element receiver antenna can still be realized by installing two dual-polarized antennas with enough spacing in between). Any performance improvement contributed by a further increase in the number of receiving antenna elements shall be viewed as an abstract representation of the future advancement in cellular technology.

\subsection{Numerical Results }
\subsubsection{\textbf{Rural Scenario}}
  \begin{figure}[t]
  \centering
  \includegraphics[width=0.48\textwidth]{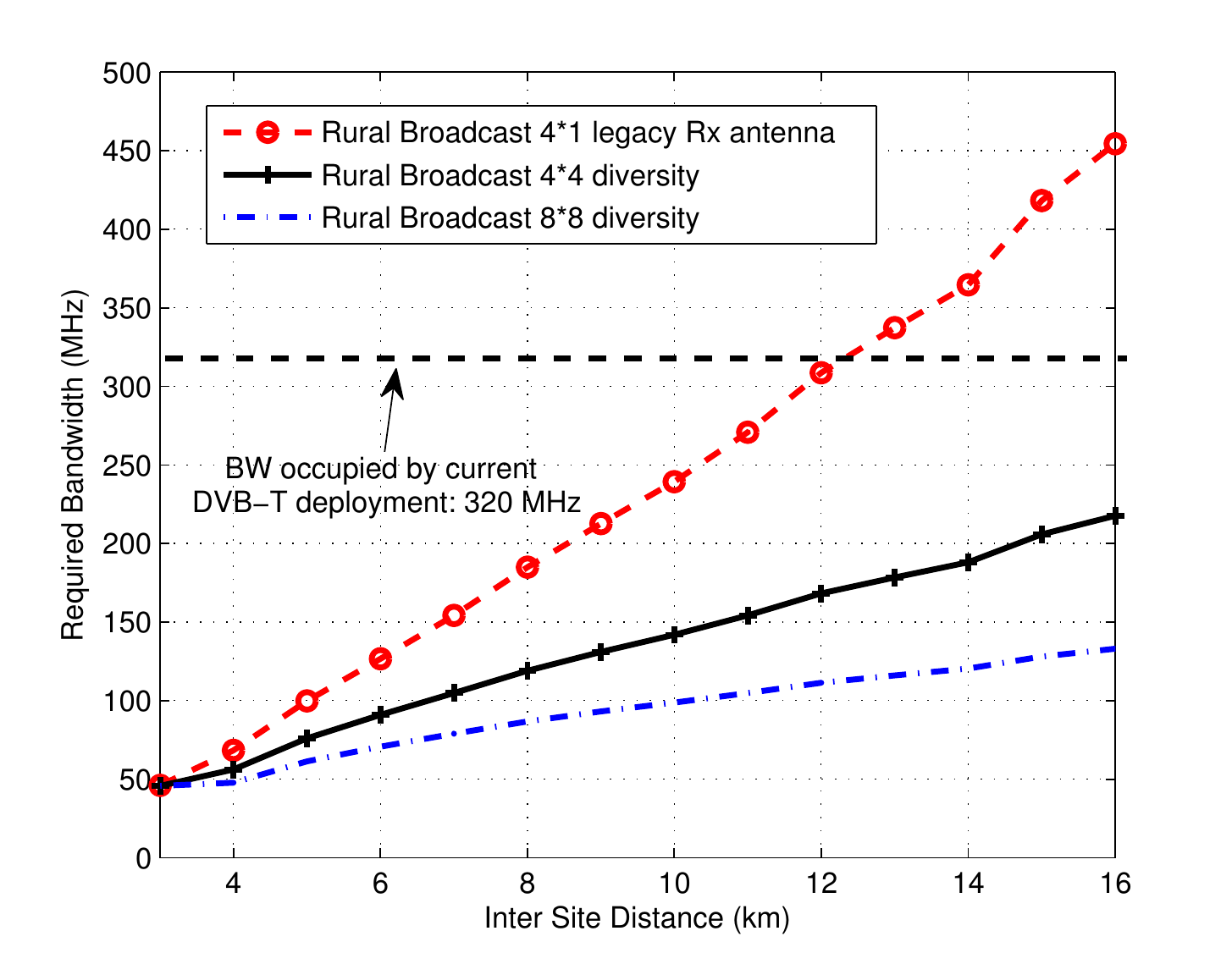}\\
  \caption{Spectrum requirement for CellTV broadcast in rural environment.}\label{fig:Broad_rural}
\end{figure}
Fig. \ref{fig:Broad_rural} depicts the spectrum requirement for the broadcast-only CellTV system with varying ISDs in rural area. It is evident that the results are sensitive to the ISD. When ISD is larger than 12 km, pure CellTV broadcast with legacy antenna cannot even be accommodated within 320 MHz. This is because the delay spreads of the received signal cannot be mitigated by the limited cyclic prefix and thus causing severer ISI than contributing to SFN gains at such large ISD. On the contrary, the effect of ISD is less noticeable in multi-antenna cases, as their diversity gains improve SINR efficiency and as such are more resilient to lower SINR caused by ISI. However, as we mentioned earlier, even with optimistic assumption on the technology advancement in 2020, the applicability of a multi-antenna receiver with 8 uncorrelated branches would be restricted because of the physical limitation of the rooftop installation and the large separation distance required for uncorrelated reception in the UHF band. Therefore, a reasonable expectation of spectrum saving for pure CellTV broadcasting is in the range of 120-160 MHz, assuming installations of new TV receiver antennas.

   \begin{figure}[t]
  \centering
  \includegraphics[width=0.48\textwidth]{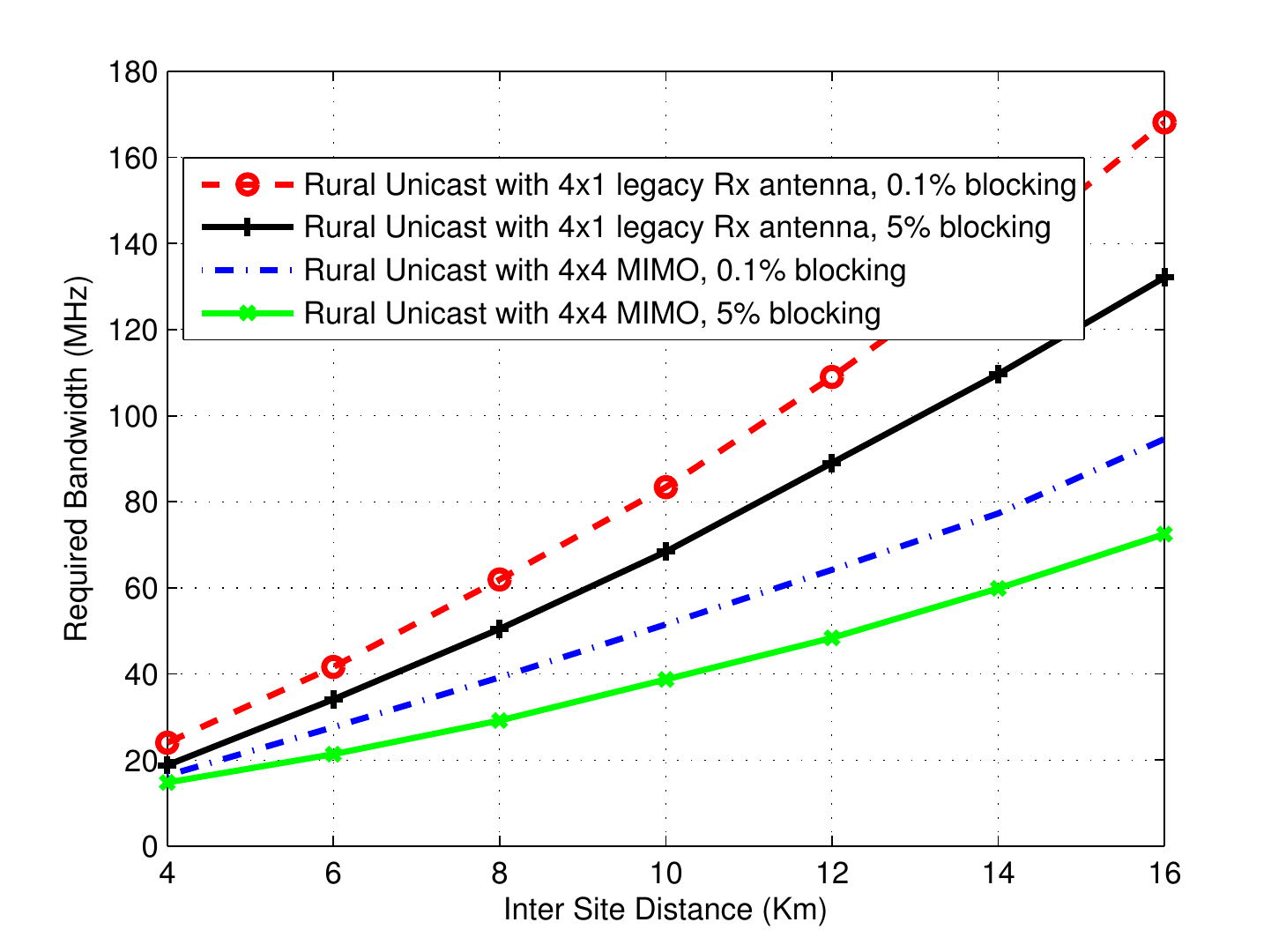}\\
  \caption{Spectrum requirement for hybrid CellTV unicast-broadcast in rural environment.}\label{fig:Hybrid_rural}
\end{figure}
The spectrum requirement for hybrid CellTV broadcast-unicast operation is illustrated in Fig. \ref{fig:Hybrid_rural}. Although the variation in ISD still has a profound impact on the spectrum demand, the amount of required spectrum is much lower than that of pure broadcast operation. Even with the legacy antenna, more than 200 MHz spectrum saving can be achieved at ISD of 12 km. The spectrum requirement can be further reduced by around $40\%$ if the receiver antenna is replaced and 4x4 MIMO is implemented. In addition, we also notice that relaxing the blocking requirement does not provide much gain in spectrum saving. Thus, it is reasonable to maintain a strict requirement on the quality of service. In general, the spectral efficiencies of the unicast links are lower than that of the SFN broadcast links. However, larger spectrum saving is still achieved by the hybrid operation due to the low population density. This is because it is far more efficient to unicast TV programs to only a few active viewers than to broadcast all the TV programs in a large cell while most of programs are not watched by anyone.

\subsubsection{\textbf{Urban Scenario}}
  \begin{figure}[t]
  \centering
  \includegraphics[width=0.48\textwidth]{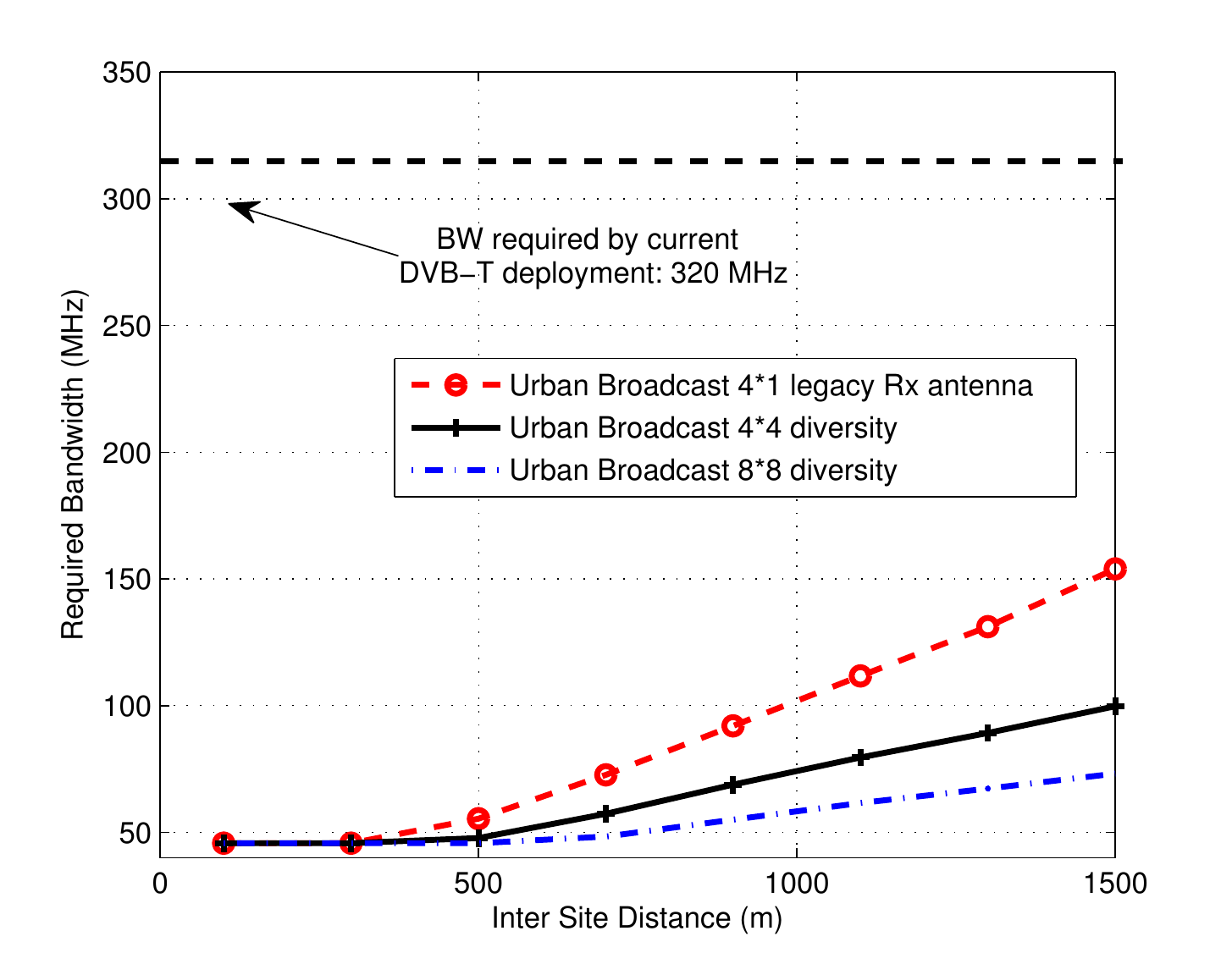}\\
  \caption{Spectrum requirement for CellTV broadcast in urban environment.}\label{fig:Broad_urban}
\end{figure}

   \begin{figure}[t]
  \centering
  \includegraphics[width=0.48\textwidth]{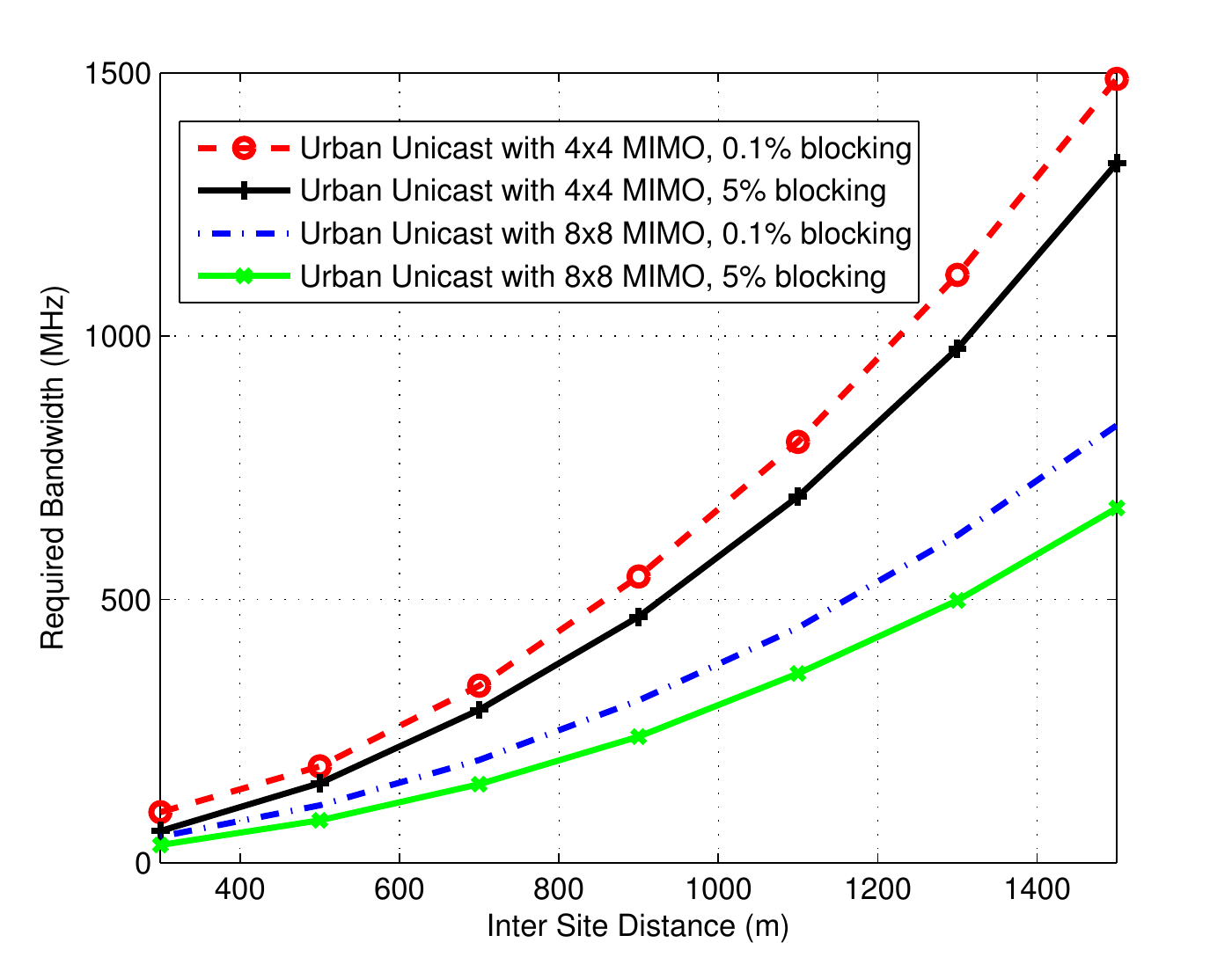}\\
  \caption{Spectrum requirement for hybrid CellTV broadcast-unicast in urban environment.}\label{fig:Hybrid_urban}
\end{figure}

The situation in the urban environment is completely the opposite of the rural scenario. Due to the higher SFN gain from a much denser cellular infrastructure in the urban area, almost 200 MHz spectrum saving can be achieved by pure broadcasting even with the legacy indoor receiver antenna (see Fig. \ref{fig:Broad_urban}). On the other hand, the hybrid broadcast-unicast operation may require more than 320 MHz spectrum to support the much higher unicast traffic in the densely populated urban areas. Particularly when the cell radius increases, a single cell will cover too many TV viewers to be supported simultaneously by the hybrid system as depicted in Fig. \ref{fig:Hybrid_urban}. Therefore, broadcast-only is considered to be the more favorable option for CellTV delivery in urban area, if we assume the TV consumption pattern remains as it is of today.

\subsubsection{\textbf{Impact of Shifting TV Consumption Pattern}}
 \begin{figure}[t]
  \centering
  \includegraphics[width=0.48\textwidth]{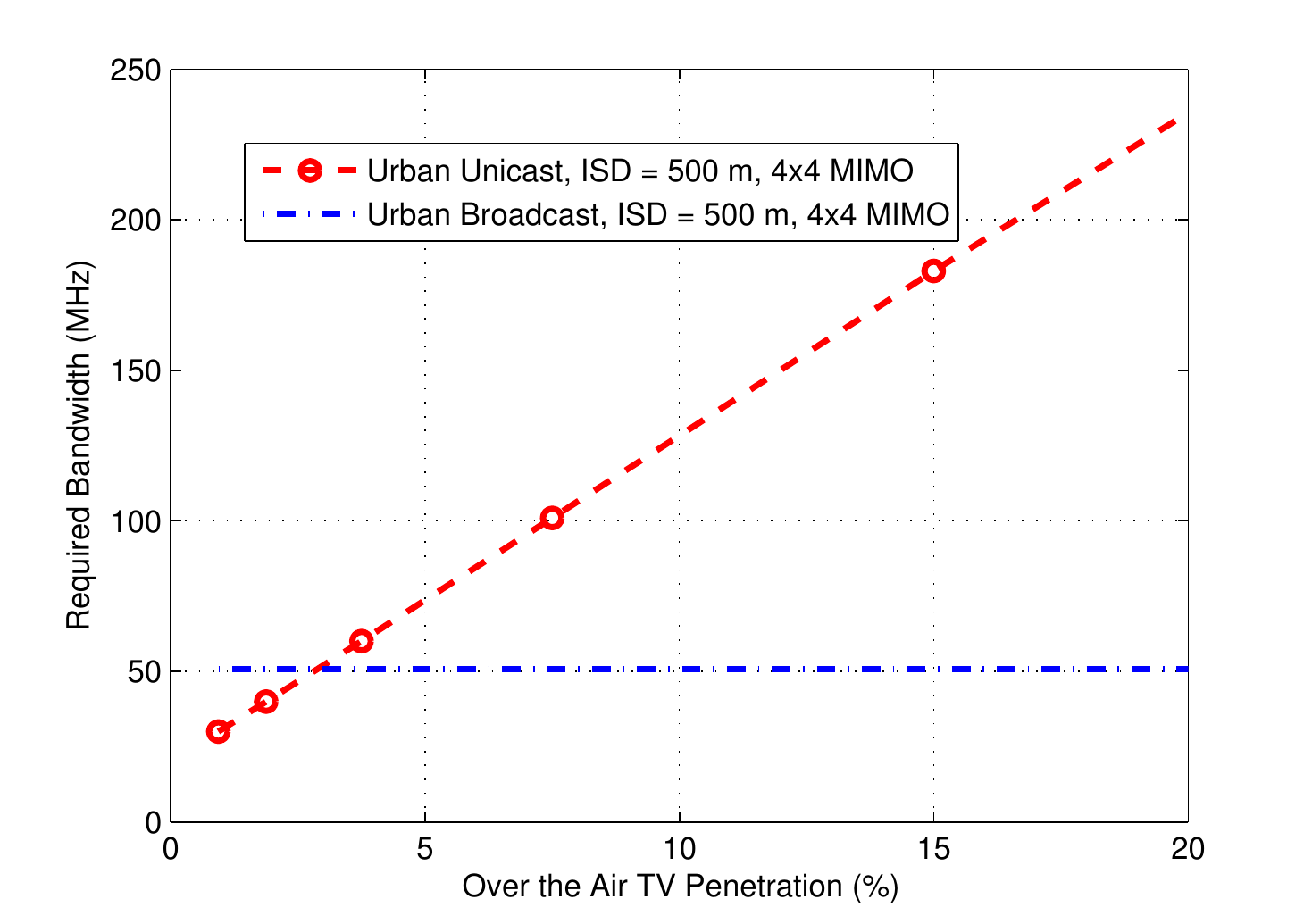}\\
  \caption{Spectrum requirement for CellTV with different terrestrial TV service penetrations in urban environment.}\label{fig:penetration}
\end{figure}	
It was observed that the hybrid operation is not beneficial in densely-populated urban areas. However, if the number of households with fixed broadband access increases and the penetration of terrestrial TV service in urban areas gradually declines, the hybrid operation with unicast capability may eventually become advantageous. We can find that the condition for hybrid operation to be more efficient than broadcasting-only in urban areas is the reduced terrestrial TV penetration from 15\% (as currently estimated) to lower than 3\% as shown in Fig. \ref{fig:penetration}).
\begin{figure}[t]
  \centering
  \includegraphics[width=0.48\textwidth]{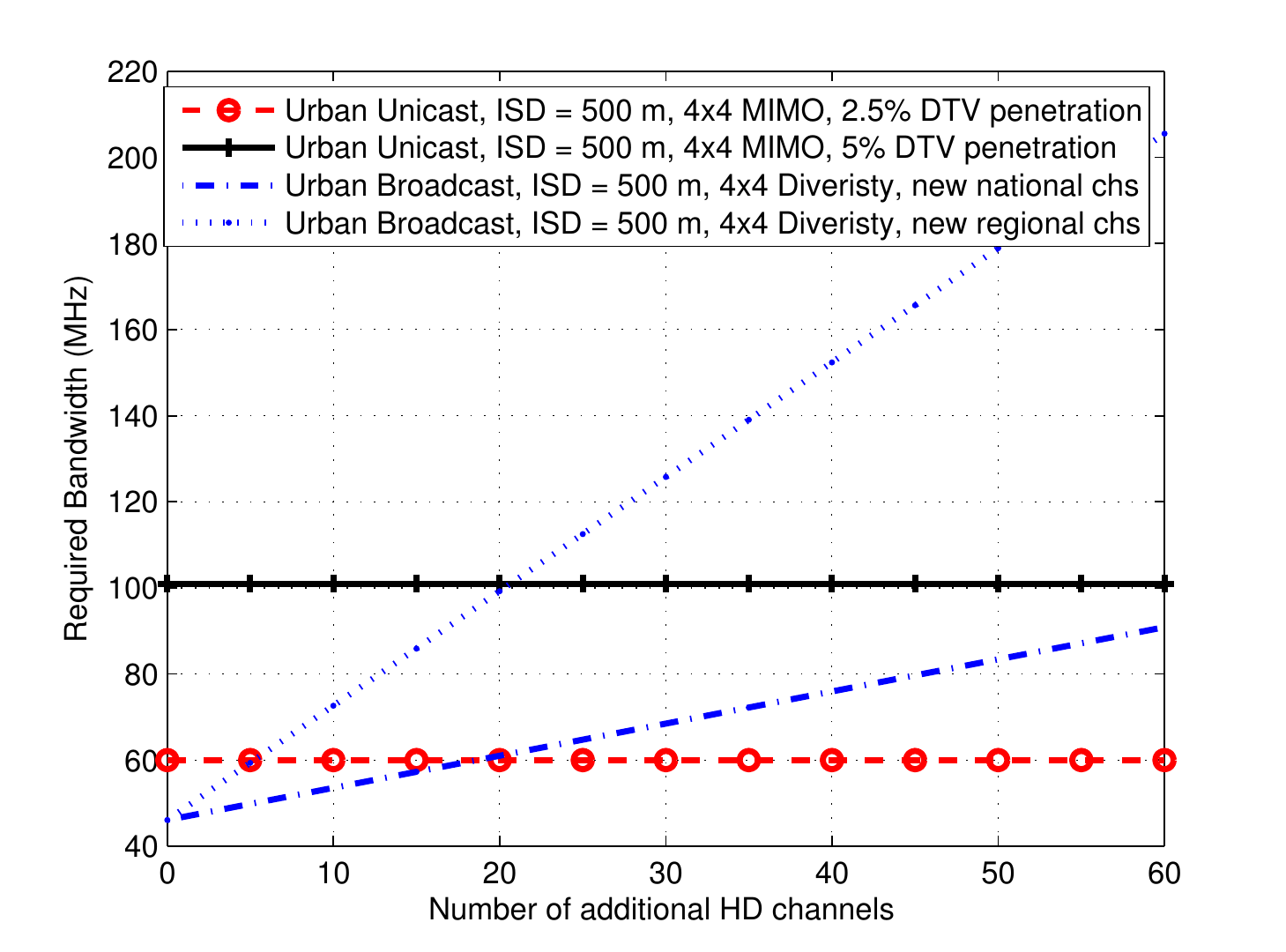}\\
  \caption{Spectrum requirement for CellTV with different number of additional TV programs in urban environment.}\label{fig:Add_ch}
\end{figure}

An advantage of the hybrid operation is VoD feature enabled by the unicast which allows new TV programs or regional contents to be easily incorporated into the existing CellTV service. Since the capacity requirement for the unicast operation only depends on the number of viewers per cell, introducing new TV programs does not require any additional spectrum or frequency re-planning. This benefit is clearly illustrated in Fig. \ref{fig:Add_ch}. Despite its advantage in spectral efficiency in dense urban areas, SFN broadcast operation would require considerably more spectrum to accommodate the increasing number of new contents. Especially if the new TV programs contain regional content, different sets of frequency channels must be used in separate SFNs. The required spectrum for broadcast-only distribution will increase drastically and become much higher than otherwise would be required for the hybrid operation.

\section{Conclusion}
\label{sec:concl}

In this paper, we have investigated the potential benefit of using cellular networks operating in 470-790 MHz as a replacement of current DTT broadcasting systems. The study targeted rural and urban Sweden in the year 2020. We have quantified the potential spectrum saving that can be achieved by this hypothetical CellTV system, using either pure broadcast over SFN or a hybrid of broadcast and unicast operations. Based on our analysis on representative Swedish rural and urban scenarios, we have reached the following major findings.

First, in rural areas CellTV only provides limited benefit when pure broadcast is considered. The spectrum saving highly depends on the performance of transceivers. The saving of 120-160 MHz is expected under reasonably optimistic assumptions about the cellular technologies with the installation of new advanced antennas at households. On the contrary, no saving at all is anticipated if some TV receivers still rely on legacy rooftop antenna. Second, in urban areas, as opposed to the rural cases, the CellTV may bring about considerable spectrum savings of up to 250 MHz without advanced TV receivers. However, the whole spectrum has to be divided for rural and urban areas to support different spectral efficiencies. It will reduce the practically achievable spectrum saving in the urban areas. Third, the feasibility of delivering TV service via unicast is dependent on the number of TV viewers per cell. In rural areas, introducing unicast can create additional spectrum saving of about 100 MHz since there are not many inhabitants. Unicast in densely populated areas is feasible, but may not be favorable compared to pure broadcasting unless the on-the-air TV penetration goes down to below $3\%$. Nonetheless, the VoD capability enabled by unicast can be regarded as the strength of CellTV.

As the penetration of DTT service and the density of cellular infrastructure differ significantly even among countries within the EU, the numerical results based on Swedish scenarios may not be directly applicable to other geographical area. However, a general conclusion from our analysis is that CellTV can be beneficial if the current trend towards more specialized programs, more local contents, and more on-demand requests continues. Mobile cellular systems, with their flexible unicast capabilities, would be an ideal platform to provide these services. Our work also shows that CellTV is not effective in replacing DTT broadcasting for the current TV viewing patterns. If the change in the TV service is modest and linear content is still the major part of the offering, then the gain would be limited. In this case, it is doubtful that the expected spectrum saving can motivate the investments in both cellular sites and TV receivers.


\bibliographystyle{IEEEtran}

\begin{IEEEbiography}{Lei Shi}
is a PhD student in the Communication Systems Department at KTH Royal Institute of Technology, Stockholm, Sweden. He joined the Radio Network lab in KTH and Center for Wireless Systems (Wireless@kth) in 2010. He received his B.S. degree in Electrical Engineering from the University of Electronic Science and Technology of China, Chengdu, China, in 2007, and his M.S. degree in Wireless Systems from KTH Royal Institute of Technology, Stockholm, Sweden, in 2009. He has participated in EU FP7 project METIS since 2012 and EU FP7 project METIS QUASAR from 2010 to 2012, focusing on spectrum sharing related topics. His research interests include radio resource management, secondary spectrum access in TV band, aggregate interference modeling and convergence of broadcast and broadband services.
\end{IEEEbiography}

\begin{IEEEbiography}{Evanny Obregon}
is currently pursuing her Ph.D. in the Communication Systems Department at KTH Royal Institute of Technology, Stockholm, Sweden.  She is also affiliated with KTH Center for Wireless Systems (Wireless@KTH). She received her B.S. degree in Electrical Engineering from the Peruvian University of Applied Science (UPC), Lima, Peru, in 2004, and her M.S. degree in Wireless Systems from KTH Royal Institute of Technology, Stockholm, Sweden, in 2009. She has participated in EU FP7 project METIS since 2012 and EU FP7 project METIS QUASAR from 2010 to 2012. She also served as a TPC member for several international conferences. Her research interests include radio resource management, dynamic spectrum access, interference modeling and future wireless architecture.
\end{IEEEbiography}

\begin{IEEEbiography}{Ki Won Sung}
(M'10) is a Docent researcher in the Communication Systems Department at KTH Royal Institute of Technology, Stockholm, Sweden. He is also affiliated with KTH Center for Wireless Systems (Wireless@kth). He received a B.S. degree in industrial management, and M.S. and Ph.D. degrees in industrial engineering from Korea Advanced Institute of Science and Technology (KAIST) in 1998, 2000, and 2005, respectively. From 2005 to 2007 he was a senior engineer in Samsung Electronics, Korea, where he participated in the development and commercialization of a mobile WiMAX system. In 2008 he was a visiting researcher at the Institute for Digital Communications, University of Edinburgh, United Kingdom. He joined KTH in 2009. He served as an assistant project coordinator of European FP7 project QUASAR. His research interests include dynamic spectrum access, energy-efficient wireless networks, cost-effective deployment and operation, and future wireless architecture.
\end{IEEEbiography}

\begin{IEEEbiography}{Jens Zander}
(S'82-M'85) is a full professor as well as co-founder and scientific director of KTH Center for Wireless Systems (Wireless@kth) at KTH Royal Institute of Technology, Stockholm, Sweden. He was past project manager of the FP7 QUASAR project assessing the technical and commercial availability of spectrum for secondary (cognitive radio) use. He is on the board of directors of the Swedish National Post and Telecom Agency (PTS) and a member of the Royal Academy of Engineering Sciences. He was the Chairman of the IEEE VT/COM Swedish Chapter (2001-2005) and TPC Chair of the IEEE Vehicular Technology Conference in 1994 and 2004 in Stockholm. He is an Associate Editor of ACM/Springer Wireless Networks Journal. His current research interests include architectures, resource and flexible spectrum management regimes, as well as economic models for future wireless infrastructures.
\end{IEEEbiography}

\begin{IEEEbiography}{Jan Bostrom}
was born in Stockholm, Sweden 1968. He received his M.Sc. degree in electrical engineering from the Royal Institute of Technology, Stockholm Sweden in 1995. He started his career as a research engineer at Ericsson working in the field of DSL (Digital Subscriber Line). In 2001 he took on the position as hardware manger and senior researcher at Repeatit AB, a company developing fixed wireless access systems.  Since 2004 he has been working at the Swedish Post and Telecom Authority where he currently holds the position of Expert Adviser mainly working with strategies for future spectrum use and spectrum regulation.
\end{IEEEbiography}

\end{document}